\documentclass[sigconf]{acmart}

\setcopyright{none}
\renewcommand\footnotetextcopyrightpermission[1]{}

\AtBeginDocument{%
  \providecommand\BibTeX{{%
    \normalfont B\kern-0.5em{\scshape i\kern-0.25em b}\kern-0.8em\TeX}}}

\setcopyright{acmcopyright}
\copyrightyear{2023}
\acmYear{2023}
\acmDOI{XXXXXXX.XXXXXXX}

\acmConference[Arxiv]{In Arxiv}{USA}

\begin{document}


\title{X-COBOL: A Dataset of COBOL Repositories}




\author{Mir Sameed Ali, Nikhil Manjunath, Sridhar Chimalakonda}
\affiliation{\textit{Research in Intelligent Software \& Human Analytics (RISHA) Lab}\\
	Department of Computer Science \& Engineering \\
	Indian Institute of Technology Tirupati
	\country{India}}
\email{{cs18b021, cs18b041, ch}@iittp.ac.in}
\authornote{Corresponding Author}


\begin{abstract}

 Despite being proposed as early as 1959, COBOL (Common Business-Oriented Language) still predominantly acts as an integral part of the majority of operations of several financial, banking, and governmental organizations. To support the inevitable modernization and maintenance of legacy systems written in COBOL, it is essential for organizations, researchers, and developers to understand the nature and source code of COBOL programs. However, to the best of our knowledge, we are unaware of any dataset that provides data on COBOL software projects, motivating the need for the dataset. Thus, to aid empirical research on comprehending COBOL in open-source repositories, we constructed a dataset of 84 COBOL repositories mined from GitHub, containing rich metadata on the development cycle of the projects. We envision that researchers can utilize our dataset to study COBOL projects' evolution, code properties and develop tools to support their development. Our dataset also provides 1255 COBOL files present inside the mined repositories. The dataset and artifacts are available at \href{https://doi.org/10.5281/zenodo.7968845}{https://doi.org/10.5281/zenodo.7968845}.  
\end{abstract}



\maketitle

\begin{figure*}[ht]
    \centering
    \includegraphics[width = \linewidth]{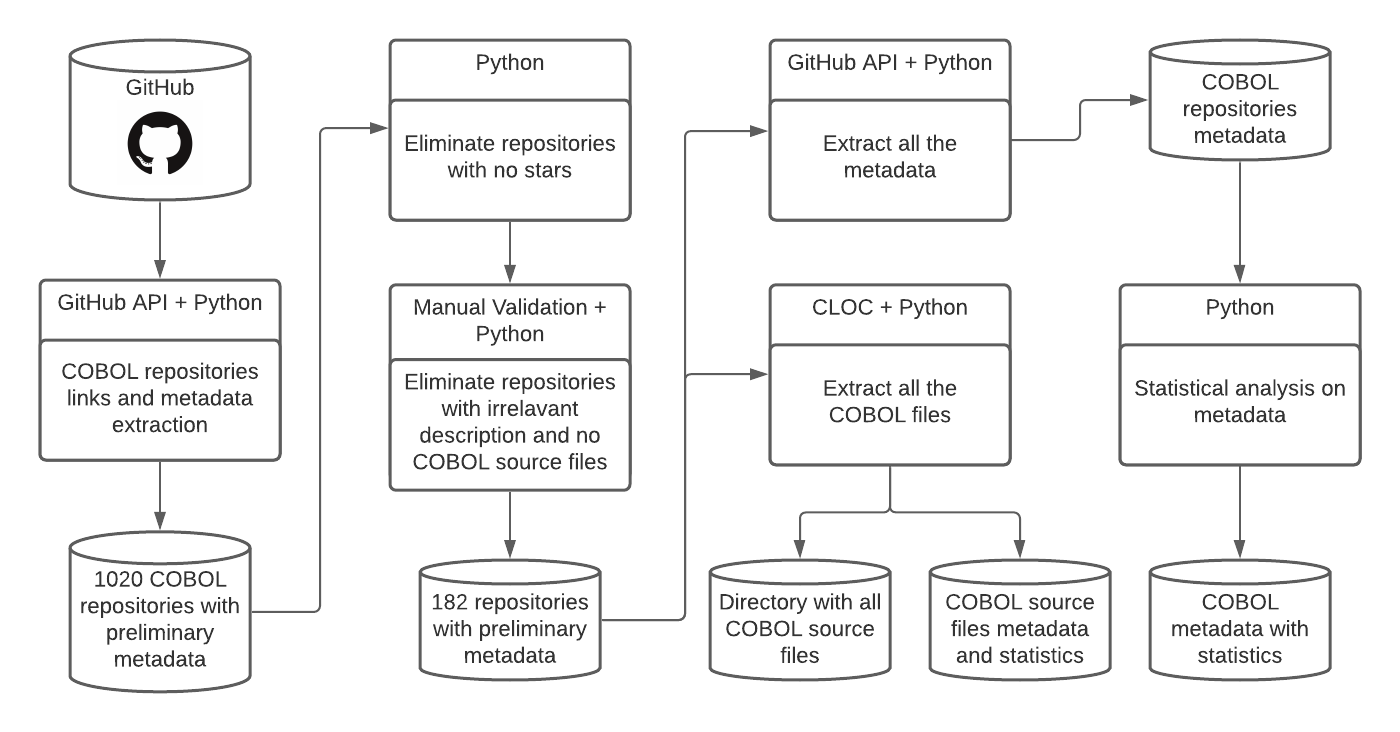}
    \caption{Data Collection Methodology}
    \label{fig:methodology}
\end{figure*}

\section{Introduction}

In StackOverflow Developer Survey 2022, programming languages such as \textit{Javascript}, \textit{Python}, \textit{Java}, and \textit{C\#} were found to be the most popular among developers. On the other hand, COBOL was the second least popular language\footnote{\url{https://survey.stackoverflow.co/2022/}}. Although COBOL is generally considered outdated, it is still being used in 80\% of financial services transactions, 95\% of ATM swipes, and there are about 220 billion lines of code, with 1.5 billion lines written every year \cite{TomTaulli}. COBOL is still processing about USD 3 trillion in commerce every day \cite{DavidCassel}. Apart from the financial sector, legacy systems written in COBOL are used in other major sectors such as healthcare and governmental institutions \cite{TomTaulli}.

Considering that legacy systems written in COBOL are still integral to vital sectors, it is essential to either maintain them or migrate them to modern systems. Migrating COBOL legacy systems to systems with modern technology is an effort-intensive work associated with an enormous amount of cost and risk. When Commonwealth Bank of Australia migrated their COBOL systems, it took them five years and cost them \$749.9 million \cite{TomTaulli}. Maintaining COBOL systems is less viable than maintaining modern systems, primarily due to a shortage of experienced COBOL programmers \cite{JohnDelaney}. Despite the enormous complications present in the maintenance and migration of  COBOL projects, software engineering research on COBOL is limited. Most efforts are towards restructuring COBOL code \cite{SELLINK2002193}, extracting knowledge and business rules \cite{Sneed}, and supporting the migration of legacy COBOL systems and industrial case studies \cite{COBOLJava,Bottom-up-COBOL,Sneed-Migration,Sneed-Migration-2}. Ciborowska et al. \cite{Contemporary-COBOL} surveyed differences in defects and defect location strategies in COBOL and modern programming languages by interviewing 30 COBOL and 74 modern programming language developers. The survey showed significant differences in defect types present in COBOL and modern programming language projects, with similar defect location strategies employed by the developers in both kinds of projects \cite{Contemporary-COBOL}. Opdebeeck et al. \cite{Opdebeeck} presented an approach to mine library usage patterns in COBOL code, which can assist the migration process.


Datasets have been created to support empirical research for different programming languages and to understand and address challenges in several software engineering areas.
Eskandani et al. constructed a dataset on open-source Serverless applications mined from GitHub \cite{wonderless}. In order to support empirical research in the domain of game engines, Vagavolu et al. presented a dataset of 526 game engine repositories mined from GitHub \cite{GE526}. While there are several datasets to support empirical research in other software engineering areas such as docker \cite{docker-Henkel, docker-schermann}, android application development \cite{android}, program equivalence \cite{EqBench}, and so on, to the best of our knowledge, there exists no dataset that caters to COBOL projects. National Computing Centre of UK provides COBOL85 test suite\footnote{\url{https://www.itl.nist.gov/div897/ctg/cobol_form.htm}}, which is a set of COBOL programs containing different features. Although the test suite can be utilized to construct COBOL parsers and compilers, it cannot support extensive analysis of the development of COBOL projects.

Hence, to facilitate the empirical research in COBOL, we present a curated dataset of 84 COBOL projects mined from GitHub, consisting of 4420 \textit{commits}, 241 \textit{pull requests}, and 727 
\textit{issues}. The resultant projects have been manually analyzed and selected based on multiple parameters. Along with the projects, we also provide 1255 COBOL program files extracted from the selected repositories.

The remainder of this paper is organized as follows. Section \ref{Data Collection} presents an overview of the data extraction methodology. Next, we present the dataset schema and the dataset statistics in Section \ref{Dataset}. Then, we describe a set of research applications of our dataset in Section \ref{Usage}. Finally, we list a few limitations of our dataset and discuss the scope of future work in Section \ref{limitations}. 

\section{Data Collection} \label{Data Collection}
This section details the data extraction process, and the implementation details are shown in Figure \ref{fig:methodology}. We have followed a similar methodology to the one that was used by Vagavolu et al. to collect a dataset on game engines \cite{GE526}. We extract all the necessary repository data required for the preliminary elimination of repositories. We then eliminate the unwanted repositories using manual evaluation and a \textit{python} script. We finally mine all the resultant repositories for metadata using the GitHub API and \textit{python}, and COBOL files using \textit{python} and \textit{CLOC}\footnote{\url{http://cloc.sourceforge.net}}.

\begin{figure*}[ht]
    \centering
    \includegraphics[width = \linewidth]{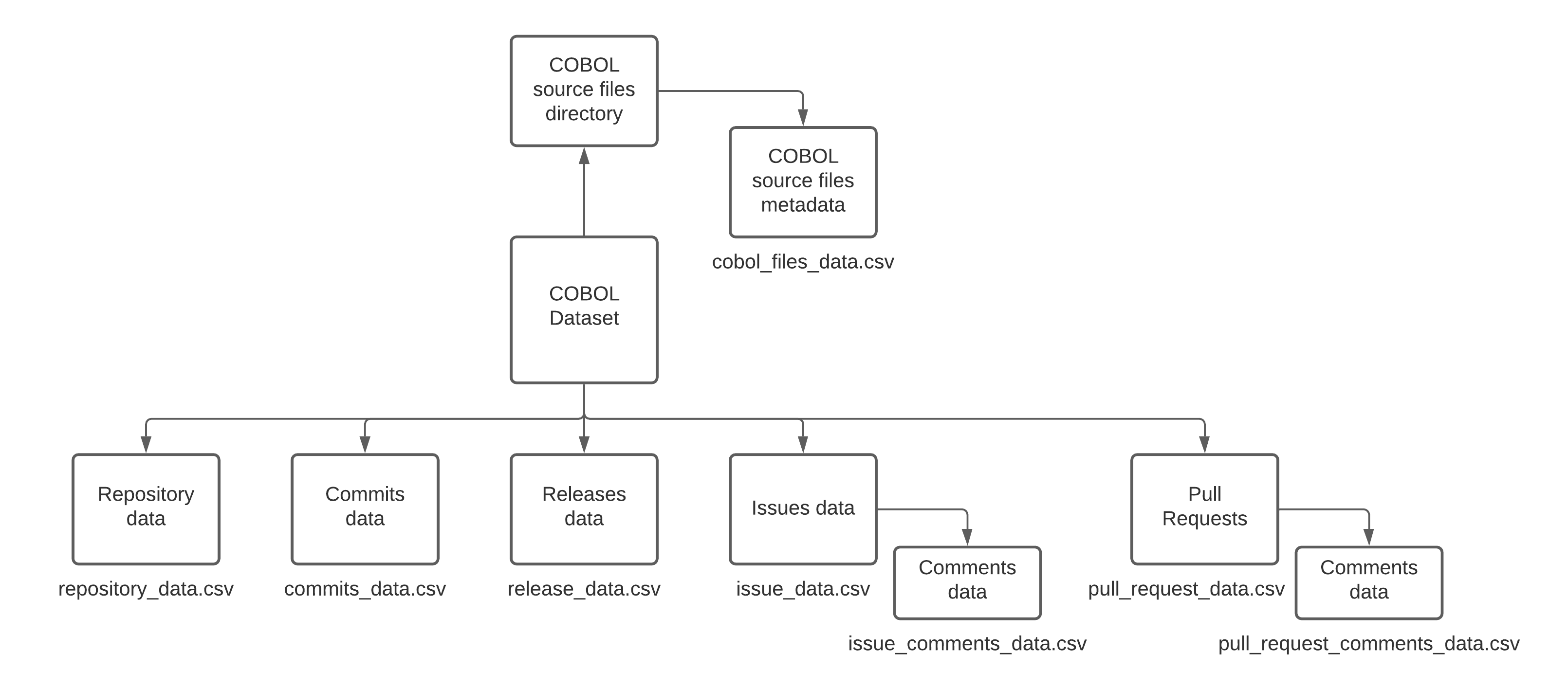}
    \caption{Dataset Schema}
    \label{fig:datasetschema}
\end{figure*}

The following sections elaborate on the dataset creation process.

\subsection{Preliminary Repository Metadata Extraction and Filtration}
We accumulated the links of all the GitHub repositories whose primary language is COBOL to be included in the dataset. We extracted the names and metadata of all these repositories using GitHub API. We ended up with 1020 repositories. In order to have a quality dataset, we eliminated all the repositories with less than one star. We also removed the repositories whose topics include the following keywords [``list'', ``course'', ``resource'', ``tutorial'', ``learning'', ``exercise'', ``example'']. At this stage, we eliminated 628 repositories and had 392 repositories to be further processed. After eliminating repositories using GitHub metrics, some repositories containing undesirable data, such as no COBOL source files and irrelevant descriptions, still existed in the dataset. One such example was \textit{SaymanNsk/Sprinter200x}, which contained compiled binaries of COBOL instead of COBOL source code. We eliminated all the false positives by a semi-automated technique validating every repository after the first elimination using the type of code and repository description. We consider the repositories related to tutorials or basic COBOL learning irrelevant for this dataset and manually filtered out 210 repositories. The final resultant repository links and names are stored in a CSV file for further data extraction. We finally have 84 repositories in the dataset.

\subsection{Extraction of metadata}
After eliminating false positives in two levels, we decided to use \textit{python} and the GitHub API to extract all the repositories' metadata using multiple GitHub access keys. For each repository, we extract metadata of 7 different categories as shown in Figure \ref{fig:datasetschema}. We also performed statistical analysis on the collected repositories calculating measures such as \textit{Max}, \textit{Mean}, and \textit{Total} for a set of metrics. The results of the anaysis are listed in Table \ref{tab:statistics}.

All the different category results were stored in separate CSV files. The number of repositories and the response for different API requests were less, which motivated us to dump data into simple CSV files instead of any database solutions. Detailed information about different metadata sections is mentioned in Database Schema.

\subsection{Extraction of COBOL source files}
We extracted all the COBOL files present in the selected 84 GitHub repositories\footnote{The dataset can be found at \url{https://doi.org/10.5281/zenodo.7968845}}. We
implemented a script that iterates over the selected repositories. During an iteration, a
repository is cloned, and \textit{CLOC} is used to identify all COBOL files present in the cloned
repository. \textit{CLOC} also provides metadata (\textit{File path},\textit{ number of Blank lines}, \textit{number
of Comment lines}, and \textit{number of Code lines}) for the identified COBOL files with the
extensions CBL, cbl, ccp, COB, cob, cobol, and cpy. All the recognized COBOL files and
the CLOC result of the cloned repository are stored in a directory named
\textit{AuthorName\_RepositoryName} in the COBOL files directory. We also retrieve the
number of commits made for each COBOL file using the GitHub API.

\section{The Dataset} \label{Dataset}

\begin{table}[]

\caption{Statistics of Metadata of the X-COBOL Dataset}
\label{tab:statistics}
\begin{tabular}{|l|l|l|l|l|}
\hline
\textbf{Metric Name}         & \textbf{Max} & \textbf{Mean} & \textbf{Total}  \\ \hline
Forks	& 40	& 4.31	& 362  \\ \hline
Size(KB)	& 203750	& 5313.13	& 446303  \\ \hline
Releases	& 12	& 0.24	& 20  \\ \hline
Open issues	& 205	& 3.39	& 285  \\ \hline
Closed issues	& 138	& 5.26	& 442  \\ \hline
Open pull requests	& 3	& 0.15	& 13  \\ \hline
Closed pull requests	& 52	& 2.71	& 228  \\ \hline
Commits	& 504	& 52.62	& 4420  \\ \hline
Commit frequency	& 29.0	& 3.41	& 286.08  \\ \hline
Committer frequency	& 2	& 0.49	& 41.18  \\ \hline
Integration frequency	& 38.0	& 3.69	& 309.73  \\ \hline
Integrator frequency	& 2	& 0.54	& 45.11  \\ \hline
Merge frequency	& 3.0	& 0.20	& 17.11  \\ \hline
Number of cobol files	& 117	& 14.94	& 1255  \\ \hline
Total comment lines	& 9091	& 500.68	& 42057  \\ \hline
Total code lines	& 23646	& 2397.37	& 201379  \\ \hline
Comment lines per cobol file	& 664	& 37.98	& 3190.16  \\ \hline
Code lines per cobol file	& 3320	& 241.38	& 20275.70  \\ \hline
Commits per cobol file	& 302	& 10.63	& 893.61  \\ \hline

\end{tabular}
\end{table}

\subsection{Dataset Schema}

The dataset contains eight CSV files, capturing different properties of the dataset, and a directory named \textit{COBOL\_Files} containing the extracted COBOL files. The overall schema demonstrating the type of data present in the eight CSV files is shown in Figure  \ref{fig:datasetschema}.

The \textit{repository\_data.csv} provides the overall activity of the repositories using metrics such as 
\textit{commit frequency}, \textit{merge frequency}, \textit{commits}, and others. \textit{repository\_data.csv} also contains repository metadata such as \textit{forks}, \textit{stars}, and \textit{size}, along with metrics capturing the overall COBOL files content present in the repository such as \textit{total COBOL files}, \textit{total blank lines}, and \textit{total code lines}. \textit{commits\_data.csv} has the data on all the commits made to the selected repositories providing information such as \textit{commit date}, \textit{commit message}, and \textit{commit author}. Information on the releases made in the selected repositories is stored in \textit{release\_data.csv}. \textit{issue\_data.csv} includes data on the issues created for the chosen repositories. The comments made on the issues are stored in \textit{issue\_comments\_data.csv}. \textit{pull\_request\_data.csv} and \textit{pull\_request\_comments\_data.csv} contain data on the pull requests made and their trailing comments, respectively. \textit{cobol\_files\_data.csv} contains the metadata on the COBOL files stored in the \textit{COBOL\_Files} directory and has information such as \textit{blank lines}, \textit{comment lines}, \textit{code lines}, and \textit{commits} made to the file. We see that this information could be potentially leveraged for analyzing COBOL projects based on varying metadata. 

\subsection{Dataset Statistics}
The dataset contains metadata and COBOL files of 84 GitHub repositories. The number of other languages used along with COBOL in the curated dataset is 61, with \textit{shell}, \textit{makefile}, and \textit{C} most used. The average number of commits made for a repository is 40, with 504 commits being the highest for \textit{debinix/openjensen}. The total number of COBOL files present in the dataset is 1255. On average, there are 21 COBOL files present in a repository. \textit{abrignoli/COBSOFT} has the maximum number of COBOL files which is 117. The average number of comments and code lines present in a single COBOL file is 729 and 1602, respectively.

\section{Potential Dataset Usage or Applications } \label{Usage}

In this section, we list some of the use cases of the X-COBOL dataset.

\subsection{Usability of COBOL Constructs}
The usability of standard constructs of a programming language is critical and has been studied in the context of programming language design. Recently, Peng et al. performed an empirical study on the usability of different features of \textit{Python} \cite{Python-Peng}. Al-Jarrah et al. conducted a similar study by analyzing 340 COBOL programs in 1979 \cite{DBLP:journals/spe/Al-JarrahT79}. We believe that an equivalent study on a more extensive set of COBOL programs can be performed using the COBOL source files present in the X-COBOL dataset.
Using the results, we can understand the usage patterns of different COBOL constructs, which can aid in optimizations in compilation and migration systems. Further, the inexperienced  COBOL developers can benefit from this study by adopting the commonly used patterns of popular COBOL constructs. 

\subsection{Understanding Bugs, Issues, Commits, Pull Requests in COBOL}

Metadata of open source projects such as \textit{bug reports}, \textit{issue reports}, \textit{commit messages}, and \textit{pull requests} have been analyzed to understand the causes and attributes of bugs present in several application environments and investigate the strategies employed to detect them \cite{catolino2019bugs,Bugs-Tensorflow}. An analogous study can be conducted using X-COBOL to comprehend the characteristics of bugs present in COBOL projects. Litecky et al. examined the types of errors and their occurrence frequencies using COBOL programs in 1976 \cite{Errors-Litecky}. This study can be extended by incorporating the project metadata such as \textit{issues}, \textit{commits}, and \textit{pull request messages} present in X-COBOL to detect the cause of bugs identified. Furthermore, techniques used to locate and rectify the identified bugs can be recognized. This analysis can benefit the COBOL developers in debugging and maintaining COBOL projects.

\subsection{COBOL Open Source Analysis}

R. van Wendel de Joode et al. \cite{1579527} have shown that the reliability of open-source software increases with an increase in software usage, with a transparent flow of information between developers and the popularity of the software. The X-COBOL dataset could be used to analyze the reliability of the open-source COBOL software in the current age, considering that COBOL is a legacy programming language. Also, following from the analysis of Choudhary et al. \cite{9054852}, it is essential to analyze the developer collaboration to measure productivity and evolution of software. Researchers can analyze these patterns in open-source COBOL software using the X-COBOL dataset. Further, this dataset can also measure the open-source community support, interest, and growth concerning a legacy programming language, COBOL.

In addition, we see immense scope for conducting empirical research on COBOL language and legacy systems in similar lines to other languages like \textit{Python}, \textit{Java} and different kinds of systems such as deep learning, games and so on. 

\section{Limitations and Future Work} \label{limitations}
We have used GitHub stars count to determine and get quality of preliminary repositories, but stargazers count, and fork count are not the only metrics to obtain a quality dataset \cite{2018}. Also, due to the manual nature of the evaluation, there might be repositories that contain data which might not be up to the intended quality. We plan to include other good quality repositories among 210 repositories eliminated and remove irrelevant repositories in the near future. Due to the small dataset size, we currently provide the dataset in different csv files and the COBOL source files as a zipped file. We plan to provide the dataset in different formats, including storing the metadata and source files in a database. Furthermore, we also intend to improve the dataset quality by evaluating the COBOL files by executing them.

\section{Conclusion} \label{conclusion}

We have presented a curated dataset containing structured metadata about the development cycle of 84 COBOL projects mined from GitHub. The dataset includes metadata on the \textit{commits}, \textit{issues}, \textit{pull requests}, and \textit{releases} of the mined repositories, along with the COBOL source files present in them. Additionally, we provide the metadata of COBOL files extracted. We expect the research community to utilize the dataset on COBOL projects to conduct empirical studies on code quality, COBOL software development practices, error analysis, security, and so on. Also, the dataset can aid in research studies and tools supporting the maintenance and migration of COBOL projects. Finally, the extracted COBOL source files can be used by researchers to perform static code analysis to develop tools that support the development of COBOL projects. 

\section{Acknowledgements}
We would like to thank Exafluence Inc. for supporting us throughout this project. 

\balance
\bibliographystyle{ACM-Reference-Format}
\bibliography{sample-authordraft}










\end{document}